\documentstyle[aps,preprint,epsf,floats,psfig,tighten]{revtex}

\newcommand{\U}{UPd$_2$Al$_3~$}
\newcommand{\vq}{\bf q\rm}
\newcommand{\vk}{\bf k\rm}

\newcommand{\vH}{\bf H\rm}
\newcommand{\vv}{\bf v\rm}

\newcommand{\la}{\langle}
\newcommand{\ra}{\rangle}

\begin{document}

\draft

\title{Gap Symmetry and Thermal Conductivity in Nodal Superconductors}

\author{P. Thalmeier$^1$ and K. Maki$^{2,3}$}

\address{$^1$Max-Planck-Institute for the Chemical Physics of Solids,
N\"othnitzer Str.40, 01187 Dresden, Germany}
\address{$^2$Max-Planck-Institute for the Physics of Complex Systems, 
N\"othnitzer Str.38, 01187 Dresden, Germany}
\address{$^3$Department of Physics and Astronomy,
University of Southern California, Los Angeles,}
\address{CA 90089-0484, USA}
\maketitle

\begin{abstract}
There are now many nodal superconductors in heavy fermion (HF)
systems, charge conjugated organic metals, high T$_c$ cuprates and
ruthenates. On the other hand only few of them have well established
$\Delta(\vk)$. We present here a study of the angular dependent
thermal conductivity in the vortex state of some of the nodal
superconductors. We hope it will help to identify the nodal directions
in $\Delta(\vk)$ of \U, UNi$_2$Al$_3$, UBe$_{13}$ and URu$_2$Si$_2$. 
\end{abstract}


Unconventional superconductivity moved to center stage in recent
years\cite{Maki01}.
After a long controversy d-wave symmetry of
both hole- doped and electron- doped high T$_c$ cuprates has now been
established \cite{vanHarlingen95,Tsuei00}. Among HF
superconductors the symmetry of $\Delta(\vk)$ has been identified only for
UPt$_3$\cite{Lussier96,Tou96} and very recently
CoCeIn$_5$\cite{Izawa01a}. As to the organic superconductors the symmetry of
$\kappa$-(ET)$_2$Cu(NCS)$_2$ has also been established recently
through the angular dependent thermal conductivity in the vortex
state\cite{Izawa01b,Won01a}. Indeed we have shown that the angular dependent
thermal conductivity provides a unique window to access the nodal
directions in $\Delta(\vk)$\cite{Won01a,Won01b,Dahm00,Won00,Won01c}. For
example from study of angular dependent thermal conductivity in
Sr$_2$RuO$_4$ in a magnetic field within the a-b plane, we conclude
that the nodal lines do not lie within the a-b plane but along the c-
axis possibly at ck$_z$=$\pm\frac{\pi}{2}$\cite{Tanatar01,Izawa01c}. Recently
a number of authors proposed that the nodal lines
in \U are not in the a-b plane (ck$_z$=0) but somewhere along the c
axis\cite{Chiao97,Jourdan99,Bernhoeft98,Sato01} (ck$_z\neq$0). The
presence of the nodal lines is clearly seen from the low temperature thermal
conductivity data \cite{Chiao97,Sun95}. Also since \U has the
hexagonal symmetry, the nodal lines should lie along the c axis. The
most obvious possibilities and their associated angular variation of
the thermal conductivity will be the focus of this work. It is
generally assumed that \U is a spin singlet superconductor since it
shows a Knight shift reduction below
T$_c$\cite{Feyerherm94,Tou95,Matsuda97}. Its magnitude is however difficult to
interpret due to contributions of the AF ordered local
moments. The excitations of the latter have been identified to mediate
the electron pairing in \U\cite{Sato01}. Theoretical analysis of this
mechanism\cite{Thalmeier01} shows that due to the crystalline electric
field anisotropy a nondegenerate odd parity state should be most
favorable. It also has a reduced spin susceptibility below T$_c$ and
an associated Knight shift. Therefore we include one odd parity nodal
gap function with ck$_z$=0 in our discussion. We want to stress that
the theory we will develop here will also help to determine the nodal
lines in UBe$_{13}$, URu$_2$Si$_2$ and UNi$_2$Al$_3$ which has
recently been proposed to exhibit odd parity spin triplet pairing
\cite{Ishida01}.\\

In the following we compute the
quasiparticle density of states and the thermal conductivity in a
vortex state at low temperature (T$\ll\Delta$) in a magnetic field
within the a-c plane. Also we limit ourselves to quasi two
dimensional systems where the Fermi surface is
approximated by a corrugated cylinder. When the ratio of Fermi velocities
$\beta$=v$_c$/v$_a$ is not too small, the angular dependent thermal
conductivity will tell the nodal position along the k$_z$- axis.\\

To calculate the quasiparticle density of states we focus ourselves to
nodal superconductors where the nodes lie at some values of
ck$_z$=$\pm\chi_0$ Also for the usefulness of our results
$\alpha$=(v$_c$/v$_a$)$^2$
should not be too small (say $\alpha >$0.1). A magnetic field $\vH$
is applied in the a-c plane with an angle $\theta$ from the c
axis. The supercurrent around vortices is circulating in the plane
perpendicular to $\vH$. Following Volovik\cite{Volovik93}, we can handle the
effect of the supercurrent within semiclassical approximation. Then the
residual density of states is given by\cite{Barash97}

\begin{equation}
g=\it Re\rm\Biggl\la\frac{C_0-ix}
{\sqrt{(C_0-ix)^2+f^2}}\Biggr\ra
\end{equation} 

where x= $|\vv\cdot\vq|/\Delta$ and f= $\Delta(\vk)/\Delta$,
C$_0= \lim_{\omega\rightarrow 0}Im(\tilde{\omega}/\Delta)$,
$\tilde{\omega}$ is the renormalized frequency and $\vk$ the
quasiparticle wave vector. Also $\vv\cdot\vq$
is the Doppler shift with $\vv$ the quasiparticle velocity and 2$\vq$
the pair momentum, the brackets in the equation above denote the
average over the vortex lattice and the Fermi surface. As to f we
limit ourselves to three simple cases with odd and even parity gap
functions respectively which are appropriate for a Fermi surface model
with cylindrical symmetry 1) f=$\sin(ck_z)$, 2) f=$\cos(ck_z)$ and 3)
f=$\cos(2ck_z)$. Then for these three cases the average is worked out
and we obtain

\begin{eqnarray}
g&=&\frac{2}{\pi}\biggl\la C_0\ln(\frac{2}{\sqrt{C_0^2+x^2}})+
x\tan^{-1}(\frac{x}{C_0})\biggr\ra\nonumber\\
&&\simeq \la x\ra +C_0(\la\ln(\frac{2}{x})\ra -1)~~
\mbox{for}~\la x\ra\gg C_0\\
&&\simeq C_0\ln(\frac{2}{C_0})+\frac{1}{2C_0}\la x^2\ra~~
\mbox{for}~\la x\ra\ll C_0\nonumber
\end{eqnarray}

We call $\la x\ra\gg$C$_0$ and  $\la x\ra\ll$C$_0$ the
superclean limit and the clean limit respectively\cite{Won01b,Won00}. Here
$\la...\ra$ means the spatial average over the vortex lattice. Now let
us assume that the quasiparticle relaxation is due to impurity
scattering in the unitarity limit. Then C$_0$ is determined by

\begin{equation}
C_0=\frac{\pi}{2}\frac{\Gamma}{\Delta}g^{-1}
\end{equation} 

Which is solved as

\begin{eqnarray}
C_0&=&\frac{\pi}{2}\frac{\Gamma}{\Delta\la x\ra}
(1+\it O\rm\frac{C_0^2}{\la x\ra^2})~~
\mbox{for}~\la x\ra\gg C_0\nonumber\\
C_0&=&(\frac{\pi}{2}\frac{\Gamma}
{\Delta}\ln(2\sqrt{\frac{2\Delta}{\pi\Gamma}}))^\frac{1}{2}
(1-\frac{\Delta}{2\pi\Gamma}\la x^2\ra)~~
\mbox{for}~\la x\ra\ll C_0
\end{eqnarray} 

Finally the residual density of states is given by

\begin{eqnarray}
g&=&\la x\ra+\frac{\pi}{2}\frac{\Gamma}{\Delta\la x\ra}
\bigl(\bigl\la\ln\frac{2}{x}\bigr\ra -1\bigr)~~
\mbox{for}~\la x\ra\gg C_0\nonumber\\
g&=&g_0(1+\frac{\Delta}{2\pi\Gamma}\la x^2\ra)~~
\mbox{for}~\la x\ra\ll C_0
\end{eqnarray}

and

\begin{eqnarray}
g_0&=&\frac{\pi}{2}\frac{\Gamma}{\Delta}C_0^{-1}\simeq
(\frac{\pi}{2}\frac{\Gamma}{\Delta}
[\ln(2\sqrt{\frac{2\Delta}{\pi\Gamma}})]^{-1})^\frac{1}{2}\nonumber\\
\la x\ra&=&\frac{2}{\pi}v_a\sqrt{eH}I(\theta)\\
\la x^2\ra&=&\frac{1}{4}v_a^2(eH)F_c(\theta)
\ln(\frac{2\Delta}{v_a\sqrt{eH}})\nonumber
\end{eqnarray}

with

\begin{eqnarray}
I(\theta)&=&(\cos^2\theta+\alpha\sin^2\theta)^{\frac{1}{4}}
\frac{1}{\pi}\int_{0}^{\pi}d\phi\nonumber\\
&&[\cos^2\theta+\sin^2\theta(\sin^2\phi+\alpha\sin^2\chi_0)\nonumber\\
&&+\sqrt{\alpha}\sin\chi_0\cos\phi\sin(2\theta)]^\frac{1}{2}\\
F_c(\theta)&=&(\cos^2\theta+\alpha\sin^2\theta)^{\frac{1}{2}}
[\cos^2\theta+(\frac{1}{2}+\alpha\sin^2\chi_0)\sin^2\theta]\nonumber
\end{eqnarray}

where $\alpha$=(v$_c$/v$_a$)$^2$ and $\chi_0$= 0, $\frac{\pi}{2}$ and
$\frac{\pi}{4}$ for f=$\sin(ck_z), \cos(ck_z)$ and $\cos(2ck_z)$
respectively. The $\theta$ dependence of F$_c(\theta)$ and
G$_c(\theta)$=I$^2$($\theta$) is shown as for $\chi$=0,
$\frac{\pi}{2}$ and $\frac{\pi}{4}$ in the isotropic case
$\alpha$=1 in Figs. 1 and 2 (dashed lines). Boundary values for
$\chi_0$=0 are given by
F$_c(\frac{\pi}{2})=\frac{1}{2}\sqrt{\alpha}=\frac{1}{2}
(\frac{v_c}{v_a})$ and
G$_c(\frac{\pi}{2})=\frac{4}{\pi^2}\sqrt{\alpha}$. Similar curves for the
anisotropic case with $\alpha$=0.6 are shown in Figs. 4 and 5
respectively. As noted elsewhere the $\theta$ dependent residual
density of states is accessible to the specific heat, the superfluid density
and the spin susceptibility\cite{Won01b,Won00}. For example 

\begin{figure}
\centerline{\psfig{figure=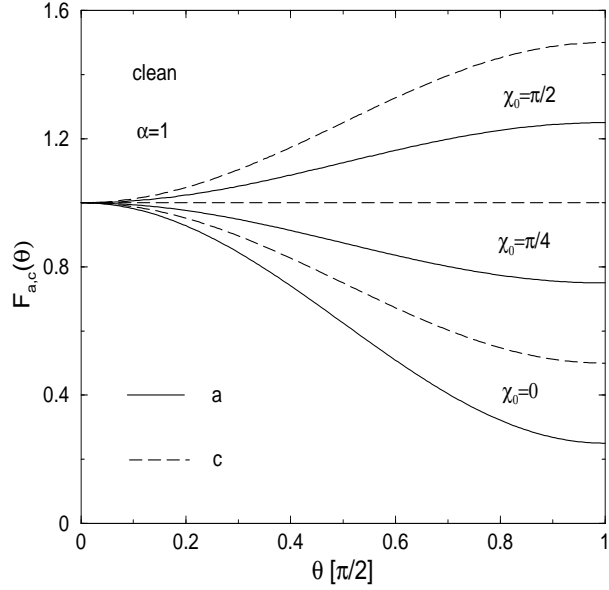,height=8cm,width=8cm}}
\vspace{1cm}
\caption{
Angular functions F$_a$($\theta$) and F$_c$($\theta$) in the clean
limit and isotropic case ($\alpha$=1) for different node line
positions $\chi_0$=ck$_z$.}
\end{figure}

\begin{figure}
\centerline{\psfig{figure=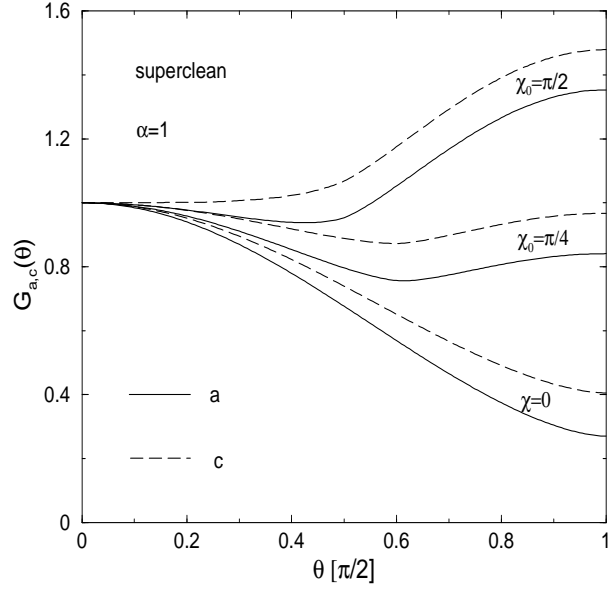,height=8cm,width=8cm}}
\vspace{1cm}
\caption{
Angular functions G$_a$($\theta$)= I($\theta$)$\tilde{I}$($\theta$) and  
G$_c$($\theta$)= I$(\theta)^2$ in the superclean limit and isotropic case
($\alpha$=1) for different node line positions $\chi_0$=ck$_z$.}
\end{figure}

\begin{eqnarray}
C_s/\gamma_NT&=&g\nonumber\\
\chi_s/\chi_N&=&g\\
\rho_s/\rho_s(H=0)&=&1-g\nonumber
\end{eqnarray} 

where C$_s$, $\chi_s$ and $\rho_s$ are the electronic
specific heat, spin susceptibility and superfluid density in the
vortex state respectively. Furthermore $\gamma_N$ is the Sommerfeld
coefficient and $\chi_N$ is the Pauli susceptibility in the normal state.

Explicitly the field dependent specific heat is given by

\begin{eqnarray}
C_s/\gamma_NT&=&\frac{2}{\pi}\frac{v_a\sqrt{eH}}{\Delta}I(\theta)\nonumber\\
&&\mbox{for the superclean limit}\\
&=&(\frac{2\Gamma}{\pi\Delta})^\frac{1}{2}
[\ln(4\sqrt{\frac{2\Delta}{\pi\Gamma}})]^\frac{1}{2}
(1+\frac{v_a^2(eH)}{8\pi\Gamma\Delta}\ln(\frac{\Delta}{v_a\sqrt{eH}})
F_c(\theta))\nonumber\\
&&\mbox{for the clean limit}\nonumber
\end{eqnarray}

In the latter the C$_s$($\theta$) angle dependence is given directly
by F$_c$($\theta$) shown in Figs. 1 and 4. For the superclean limit it
is given by I($\theta$)=G$_c$($\theta$)$^\frac{1}{2}$ shown in Fig. 3.

\begin{figure}
\centerline{\psfig{figure=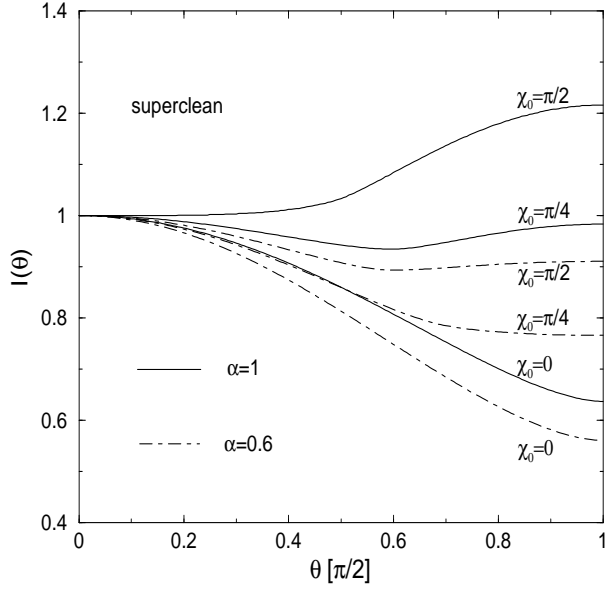,height=8cm,width=8cm}}
\vspace{1cm}
\caption{
Angular function I$(\theta)$=G$_c$($\theta$)$^\frac{1}{2}$ $\sim$
g$(\theta)$ which determines the specific heat in the superclean limit for
$\alpha$=1 and 0.6 and different node line positions $\chi_0$=ck$_z$.}
\end{figure}

We can now calculate the thermal conductivity within a similar
approximation for T$\ll\Delta$ and obtain

\begin{eqnarray}
\frac{\kappa_c}{\kappa_{nc}}&=&
\frac{2}{\pi}\frac{v_a^2(eH)}{\Delta^2}I^2(\theta)\nonumber\\
&&\mbox{for the superclean limit}\\
\frac{\kappa_c}{\kappa_{0c}}&=&
1+\frac{v_a^2(eH)}{6\pi\Gamma\Delta}F_c(\theta)
\ln(2\sqrt{\frac{2\Delta}{\pi\Gamma}})\ln(\frac{2\Delta}{v_a\sqrt{eH}})
\nonumber\\
&&\mbox{for the clean limit}\nonumber
\end{eqnarray}

where $\kappa_{nc}$ is the thermal conductivity in the normal state
along the c axis while $\kappa_{0c}$=$\kappa_{c}$(H=0) in the clean
limit. The linear H dependence of the heat current is
the signature of the superclean limit. In the clean limit, on the
other hand, the thermal conductivity starts from the nonvanishing
value for H=0. Also the H dependence is sublinear in H. Similarly the
heat current parallel to the a axis is given by 

\begin{eqnarray}
\frac{\kappa_a}{\kappa_{na}}&=&
\frac{2}{\pi}\frac{v_a^2(eH)}{\Delta^2}I(\theta)\tilde{I}(\theta)\nonumber\\
&&\mbox{for the superclean limit}\\
\frac{\kappa_a}{\kappa_{0a}}&=&
1+\frac{v_a^2(eH)}{6\pi\Gamma\Delta}F_a(\theta)
\ln(2\sqrt{\frac{2\Delta}{\pi\Gamma}})\ln(\frac{2\Delta}{v_a\sqrt{eH}})
\nonumber\\
&&\mbox{for the clean limit}\nonumber
\end{eqnarray}

where

\begin{eqnarray}
\tilde{I}(\theta)&=&(\cos^2\theta+\alpha\sin^2\theta)^{\frac{1}{4}}
\frac{1}{\pi}\int_{0}^{\pi}d\phi(1+\cos(2\phi))\nonumber\\
&&[\cos^2\theta+\sin^2\theta(\sin^2\phi+\alpha\sin^2\chi_0)\nonumber\\
&&+\sqrt{\alpha}\sin\chi_0\cos\phi\sin(2\theta)]^\frac{1}{2}\\
F_a(\theta)&=&(\cos^2\theta+\alpha\sin^2\theta)^{\frac{1}{2}}
[\cos^2\theta+(\frac{1}{4}+\alpha\sin^2\chi_0)\sin^2\theta]\nonumber
\end{eqnarray}

\begin{figure}
\centerline{\psfig{figure=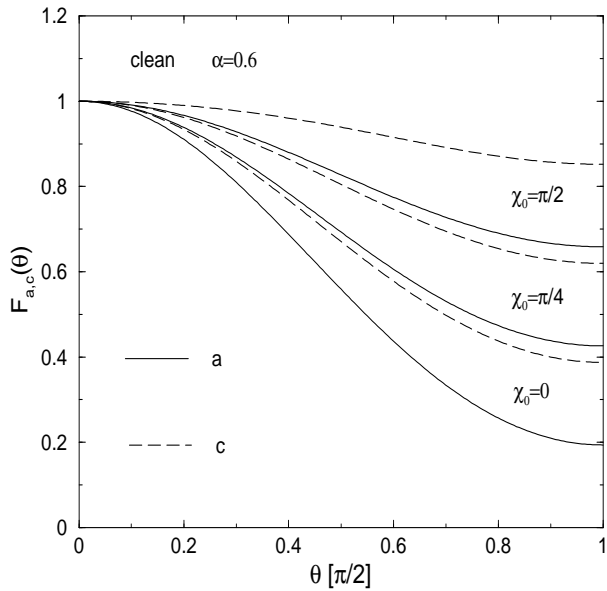,height=8cm,width=8cm}}
\vspace{1cm}
\caption{
Angular functions F$_a$($\theta$) and F$_c$($\theta$) in the clean
limit and anisotropic case ($\alpha$= 0.6) for different node line
positions $\chi_0$=ck$_z$.}
\end{figure}

The $\theta$- dependence of F$_a(\theta)$ and G$_a(\theta)$=
$I(\theta)\tilde{I}(\theta)$ for $\alpha$=1 and $\chi$=0,
$\frac{\pi}{2}$ and $\frac{\pi}{4}$ are shown in Figs. 1 and 2 (full
lines) and similar curves for $\alpha$=0.6 in Figs. 4 and 5. The
boundary values for $\chi_0$=0 are
F$_a(\frac{\pi}{2})=\frac{1}{4}\sqrt{\alpha}$ and
G$_a(\frac{\pi}{2})=\frac{8}{3\pi^2}\sqrt{\alpha}$. From the
superclean limit in both cases a and c in Figs. 2
and 5 we notice that the angular dependence is decidely nonmonotonic for the
even gap function as opposed to the monotonic decrease of G$_{a,c}$
($\chi_0\neq$0) in the odd parity case ($\chi_0$=0). Therefore thermal
conductivity measurements in \U should be able to decide definitely
between the two cases. So far we have neglected the phononic thermal
conductivity, which may be easily considered as in \cite{Won01a}.

Comparison of $\alpha$=1 and $\alpha$=0.6 results show that the
difference in $\theta$- dependent curves F$_{a,c}(\theta)$ and
G$_{a,c}(\theta)$ for the various node positions $\chi_0$ becomes
smaller with decreasing $\alpha$. For specific compounds the value of
$\alpha$ may be obtained from the upper critical field anisotropy
according to v$_c$/v$_a$=H$_{c2\parallel}$/H$_{c2\perp}$. For a small
value $\alpha$=0.1 the difference at $\theta=\frac{\pi}{2}$ (for the
various $\chi_0$) becomes also quite small ($\sim$10\%). This means that
in the very anisotropic cases like Sr$_2$RuO$_4$
($\sqrt{\alpha}=\frac{1}{20}$, i.e. $\alpha$=0.0025) the angular
$\theta$ dependence therefore can only determine the proper (ab-) plane of
node lines\cite{Izawa01c} but not the position along the c- axis. On
the other hand \U with $\alpha$= 0.69 is a much more favorable case
and the $\theta$- dependence of $\kappa_{a,c}$, C$_s$ should render also the
node position along the c- axis. 
 
\begin{figure}
\centerline{\psfig{figure=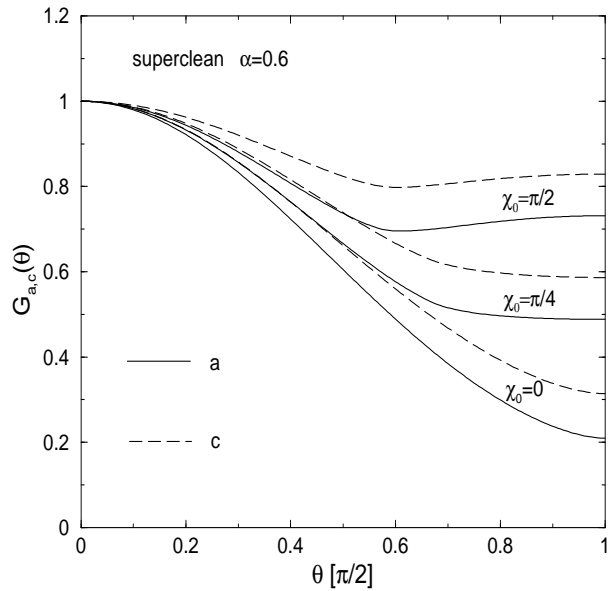,height=8cm,width=8cm}}
\vspace{1cm}
\caption{
Angular functions G$_a$($\theta$)= I($\theta$)$\tilde{I}$($\theta$) and  
G$_c$($\theta$)= I$(\theta)^2$ in the superclean limit and anisotropic case
($\alpha$=0.6) for different node line positions $\chi_0$=ck$_z$.}
\end{figure}

In conclusion there are many experimental techniques to discover the nodal
superconductivity. In particular the $\sqrt{H}$- dependence of the
specific heat provides the clearest signal for
it\cite{Won01b,Volovik93}. On the other hand to determine the nodal
directions in $\Delta(\vk)$ we need
more delicate investigations like the phase sensitive
experiments\cite{vanHarlingen95,Tsuei00}. However, the elegant
tricrystal experiments appear to not transport to other nodal
superconductors aside from high T$_c$ superconductors. In this
circumstance the angular dependent
thermal conductivity will provide a unique window to look at the nodal
directions\cite{Won01a,Won01b}. From the angular dependence of the thermal
conductivity Izawa et al have succeeded in deducing the symmetry of
$\Delta(\vk)$ in Sr$_2$RuO$_4$\cite{Izawa01c},
CeCoIn$_5$\cite{Izawa01a} and more recently
$\kappa$-(ET)$_2$Cu(NCS)$_2$\cite{Izawa01b}. Therefore it is of
great interest to explore this technique to still unidentified order
parameter in \U ,UNi$_2$Al$_3$ and other nodal superconductors.\\

\noindent
{\em Acknowledgement}\\ 
We thank Yuji Matsuda for useful discussion on the thermal
conductivity in the nodal superconductors and Yoh Kohori for
indicating us Refs. \cite{Tou95} and \cite{Matsuda97}

\end{document}